\documentclass[reprint, amssymb, showpacs, preprintnumbers, aps, sort&compress, numbers]{revtex4-2}
\usepackage{amsmath}                   
\newcommand{\vect}[1]{\boldsymbol{#1}} 

\usepackage{graphicx}                  
\usepackage{etoolbox}                  
\usepackage[dvipsnames]{xcolor}
\mathchardef\mhyphen="2D 
\usepackage[utf8]{inputenc}
\DeclareUnicodeCharacter{2212}{-}

\makeatletter
\newcommand*{\rom}[1]{\expandafter\@slowromancap\romannumeral #1@}
\makeatother

\begin{document}


\title{Experimental studies on collision between self-propelling liquid crystal droplets in quasi-two-dimensional geometry}
\author{Swarnak Ray}
\author{Arun Roy}
\email{aroy@rri.res.in}
\affiliation{%
 Soft Condensed Matter Group, Raman Research Institute, Bangalore 560080, India
}%


\date{\today}

\begin{abstract}
 Active droplets undergoing micellar solubilization can self-propel themselves by emitting filled micelles from their surface and are by nature anti-chemotactic. These droplets get repelled by their own trail of filled micelles or by other neighbouring droplets. We experimentally study the various types of scattering events between two such active droplets. We define different collision parameters to characterize the scattering events and monitor the time evolution of these parameters about the closest approach between the droplets.  For generic oblique collisions of the droplets, we find four major types of scattering events in our experimental studies depending on the initial angle of approach, the initial Peclet numbers, the delay time, and the initial distance between the droplets. For higher Peclet number droplets, the existence of a bound state was also found.
\end{abstract}
\maketitle



\section{introduction}

Active matter consisting of very many self-moving components has recently attracted the interest of researchers.  Nature has developed numerous strategies to allow living species to explore and conquer their environment. As a matter of fact, the larger the being, the more complex the strategy. Very simple organisms, such as prokaryotic cells, have developed efficient propulsion mechanisms despite their lack
of complex structure. Some biological systems of interest are bacterial suspensions, cell layers, terrestrial, aquatic, and aerial flocks \cite{Mehes2014collective, Schnyder2017collective, Giardina2008collective, Pavlov2000, Gueron1996}.

In recent years there has been a growing interest in developing biomimetic artificial micron-sized swimmers such as colloidal Janus particles \cite{Moran2017phoretic}. Another typical example of an artificial microswimmer is a liquid droplet swimming in an immiscible fluid due to self-generated Marangoni flow. This requires an asymmetry of surface tension at the interface of the droplet. This asymmetry in surface tension can be maintained by a non-uniform distribution of surfactants on the interface.  Such non-uniform distribution of surfactant molecules on the interface can be generated by chemical reactions \cite{Banno2012ph, Ban2013ph, Kitahata2011sp, Kasuo2019, Suematsu2021, Suematsu2016, Suematsu2019, Thutupalli2011, Thutupalli2013tuning, Schmitt2013, Yoshinaga2012} or by micellar solubilization. In the micellar solubilization process, a droplet of one fluid slowly dissolves in a micellar surfactant solution of another fluid by forming filled micelles. The filled micelles are formed by acquiring some molecules of the dissolving inner fluid of the droplet into the core of the micelles which minimizes the overall free energy of the system. The evidence for the emission of filled micelles has been observed using fluorescence microscopy studies. It was reported that the droplets leave behind a trail of filled micelles \cite{Jin2017, Hokmabad2021} along the trajectory during their self-propulsion.

  Earliest reports of self-generated flow fields by such solubilizing droplets can be found in \cite{Chen1997rates, Chen1998rates, Pena2006, Peddireddy2012}. In recent years, there have been further studies on droplet motion due to micellar solubilization \cite{Izri2014, Herminghaus2014, Krueger2016Curling, Jin2017, Moerman2017, Yamamoto2017, Suga2018, Izzet2020, Suda2021, Dwivedi2021, Hokmabad2021, Castonguay2023, Ray2023Experimental}. Some of these systems involve water droplets suspended in solutions of a non-ionic surfactant in an organic oil \cite{Izri2014, Suda2021, deBlois2021}, while some others involve oil droplets suspended in aqueous solutions of ionic surfactants \cite{Herminghaus2014, Krueger2016Dimensionality, Jin2017, Moerman2017, Yamamoto2017, Suga2018, Izzet2020, Dwivedi2021, Hokmabad2021}. Though the solubilization process starts above the critical micellar concentration (CMC), it has been found that self-propulsion of the droplet happens only above a threshold total concentration of the surfactant in the outer fluid which is much greater than CMC. The self-propulsion of droplets with the inner fluid in the isotropic \cite{Izri2014, Moerman2017,deBlois2019, Izzet2020, Jin2021, deBlois2021}, nematic \cite{Herminghaus2014, Krueger2016Curling, Jin2017, Suga2018, Dwivedi2021} or cholesteric phases \cite{Yamamoto2017, Yamamoto2019} has been reported. Several mathematical models aiming to explain the mechanism behind self-propulsion have also been proposed in the literature \cite{Herminghaus2014, Morozov2019nonlinear, Izzet2020, Morozov2020, Li2022, Ray2023}.

In chemotaxis, cells, and microorganisms respond to the presence of certain chemicals and move toward or away from them. Similar behavior can also be observed in artificial systems such as self-propelled droplets undergoing micellar solubilization. The interaction between two such active particles is of considerable interest. The complexity of the living systems often makes it difficult to study such interactions in a controlled fashion. The artificial self-propelling droplets can be exploited to study such phenomena more systematically. Moerman \textit{et al.} studied the interaction between two such solubilizing droplets below the instability threshold for self-propulsion \cite{Moerman2017}. The collective behavior of active droplets in different levels of confinement has also been studied in recent years \cite{Thutupalli2018, Krueger2016Dimensionality, Hokmabad2022}. Meredith \textit{et al.} showed how chemotactic signaling between microscale droplets of different fluids in micellar solutions can result in predator–prey-like chasing phenomena \cite{Meredith2020}. There have also been some studies on the scattering of the droplets from the trail of a single droplet and the chemotactic effect of many such droplets  \cite{Jin2017, Hokmabad2022Chemotactic}. Recently, there has been a study on Peclet number-dependent interaction between active droplets \cite{Dwivedi2024}.

The study of two-particle interactions is the first step in understanding the collective behaviour of such droplets. In this article, we report the self-propulsion of liquid crystalline droplets suspended in an ionic micellar solution and the scattering of a pair of droplets for different droplet sizes and initial conditions. We find four typical types of scattering processes for acute and obtuse angles of approach of the droplets. We also find the existence of bound states for larger droplets.


\section{Experiment}
\label{Experiment}
We used the common liquid crystalline (LC) compound 5CB (4-pentyl-4$^\prime$-cyanobiphenyl) for our experiments. At room temperature, 5CB is in the nematic state and has a nematic to isotropic transition temperature of 308.15 K. For micellar solution, we use the ionic surfactant sodium dodecyl sulfate (SDS) in water. All chemicals were obtained from Sigma-Aldrich and used as received. The surfactant solution of the given concentration was prepared by dissolving a suitable amount of SDS in deionized water (Milli-Q) and keeping the solution overnight at room temperature. The small droplets for the collision experiments were first prepared using a tip sonicator in a 0.2 wt $\%$ SDS solution. This low concentration of the SDS enables us to stabilize the LC droplets without much solubilization. A suitable amount of this stock emulsion was then added to a 9 wt $\%$ SDS solution to obtain a dilute solution of the droplets suitable for observation under the microscope. This higher concentration of SDS in the micellar solution enables the self-propulsion of the LC droplets. Droplet trajectories were observed using a digital camera (Canon EOS 80D) attached to a polarizing optical microscope (Olympus BX51). The temperature of the sample was monitored using a hotstage (LINKHAM LTS420E) and a temperature controller (LINKHAM T95) with a temperature resolution of  0.1 K. The experiments were conducted at a fixed temperature of 313.15 K.  All of the experiments were performed in sample cells consisting of two parallel glass plates, separated by a distance much smaller than their lateral dimensions. The sample cells were prepared by adhesive spacers of thickness 120 $\mu m$ (Secure seal imaging spacer, Grace Biolabs) which allowed the bonding of two glass plates to it. The measured separation between the plates varied in the range  135 - 150 $\mu m$. The motion of the droplets was recorded using video microscopy with a 5x microscope objective.  No attempt was made to density match the outer and the inner fluids.
 
The position coordinates of the droplets in the horizontal plane were extracted by image processing using a Matlab subroutine based on Hough transformation. Small vertical excursions of the droplets during the motions were neglected. The following processing steps were used in the determination of the trajectories: Cropping $\rightarrow$ Binarization  $\rightarrow$ Gaussian blurring $\rightarrow$ Edge detection using Canny edge detector $\rightarrow$ Finding circles $\rightarrow$ tracking. The tracking of the positions was performed using the Matlab implementation by Blair and Dufresne of the Crocker and Grier tracking algorithm \cite{Crocker1996}. The trajectories of the droplets were determined using an Eilers smoother with a 3rd-order penalty \cite{Eilers2003}. 
\begin{figure*}
    \centering
    \includegraphics[width =0.9\linewidth]{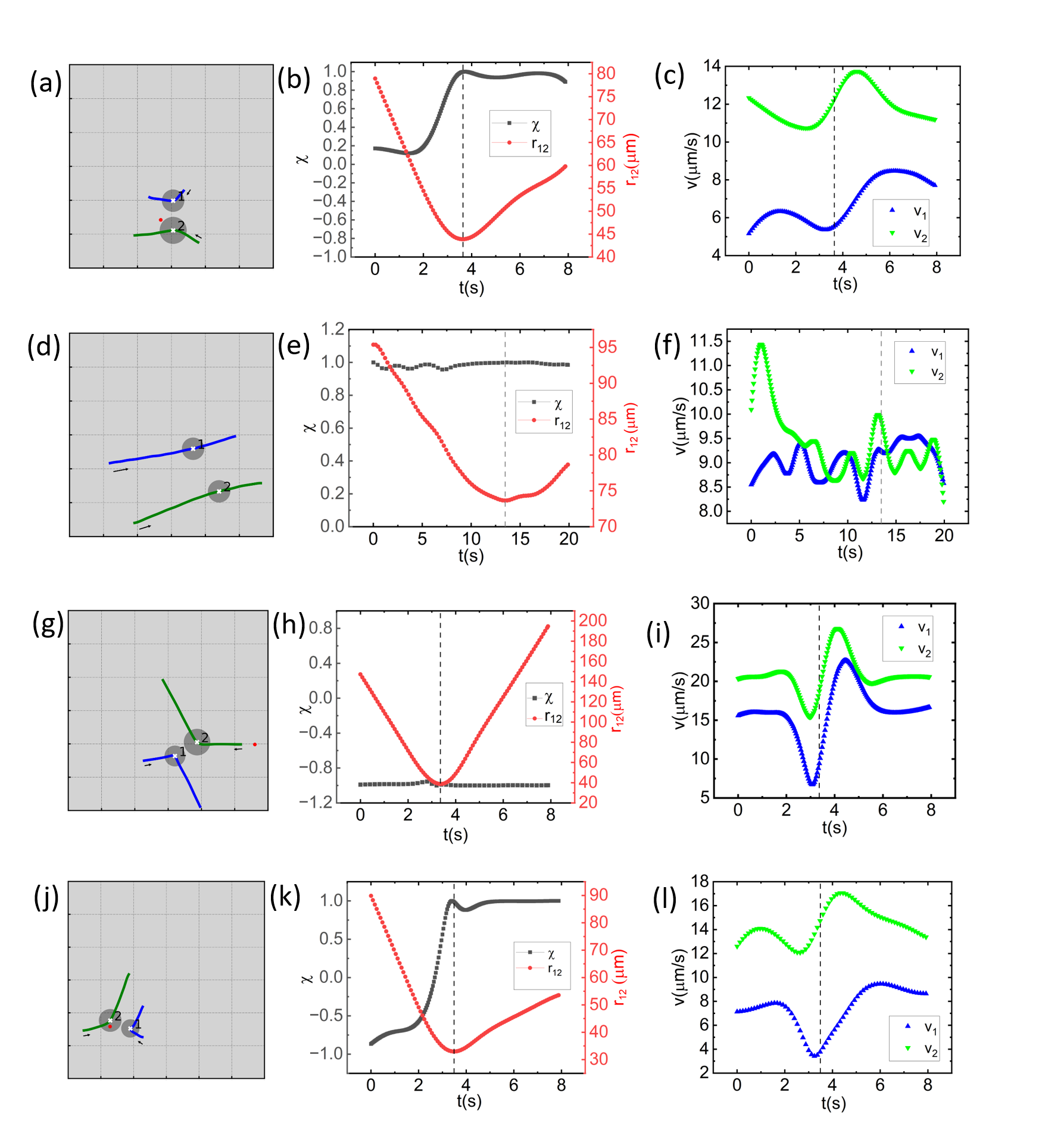}
    \caption{ Different types of collision events observed in our experiments. Acute angle of approach: (a) trajectories of the colliding droplets, (b) relative direction cosine $\chi$ and inter-droplet distance $r_{12}$, (c) instantaneous speeds $v_1, v_2$ of the droplets as a function of time during the collision for the set of initial parameter values  \{$Pe_1, Pe_2, \chi(0),r_{12}(0), \Delta \tau$\} = \{1.2, 3.6, 0.171, 79 $\mu m $, 5.11 s\}. Corresponding figures for almost parallel angle of approach: (d) trajectories, (e) $\chi$ and $r_{12}$, (f)  $v_1, v_2$ for the set of initial parameter values \{$Pe_1, Pe_2, \chi(0), r_{12}(0), \Delta \tau$\} = \{2.1, 2.4, 0.98, 95 $\mu m$, 22.35 s\}. Corresponding figures for almost antiparallel angle of approach: (g) trajectories, (h) $\chi$ and $r_{12}$, (i)  $v_1, v_2$ for the set of initial parameter values \{$Pe_1, Pe_2, \chi(0), r_{12}(0), \Delta \tau$\} = \{3.6, 5.9, -0.989, 147, 11.82 s\}. Corresponding figures for obtuse converging angle of approach: (j) trajectories  (k) $\chi$ and $r_{12}$, (l) $v_1, v_2$ for the set of initial parameter values \{$Pe_1, Pe_2, \chi(0), r_{12}(0), \Delta \tau$\} = \{1.4, 3.2, -0.866, 90, 2.86 s\}.  The grid line spacing in fig. (a), (d), (g), (j) is 50 $\times$ 50 $\mu m^2$. The dashed line corresponds to the time of closest approach.} 
    \label{fig:collision}
\end{figure*}


\begin{figure*}
    \centering
    \includegraphics[width = 0.8\linewidth]{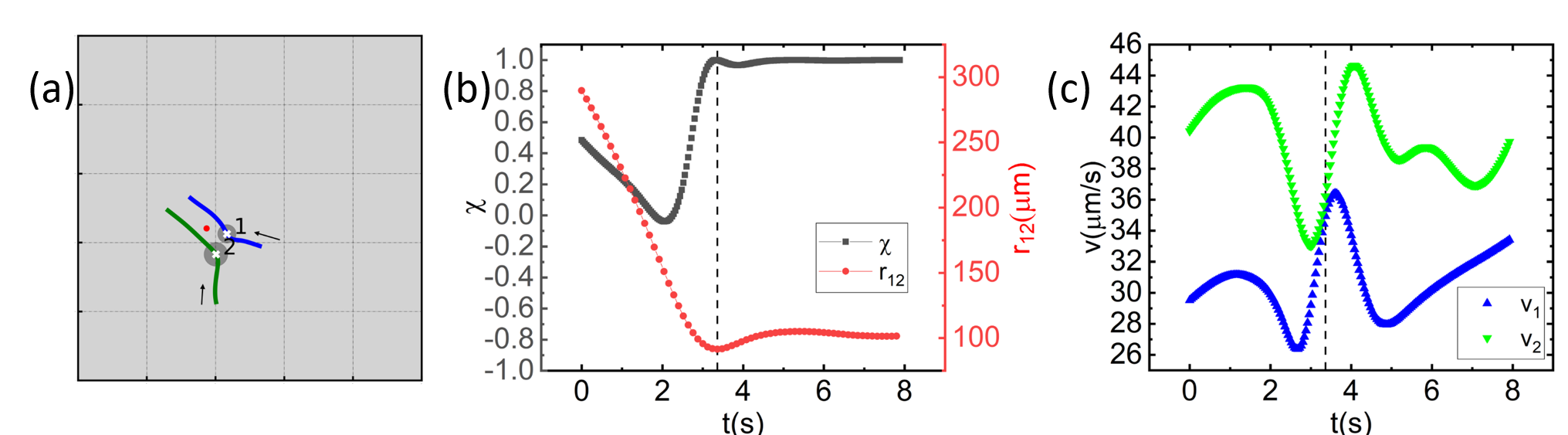}
    \caption{ Collision process showing the existence of bound states: (a) trajectories of the colliding droplets, (b) relative direction cosine $\chi$ and inter-droplet distance $r_{12}$, (c) instantaneous speeds $v_1, v_2$ of the droplets as a function of time during the collision for the set of initial parameter values  \{$Pe_1, Pe_2, \chi(0), r_{12}(0), \Delta \tau$\} = \{11.8, 20.8, 0.425, 290 $\mu m$, 0.2 s\}. The grid line spacing in fig. (a) is 200 $\times$ 200 $\mu m^2$. The dashed line corresponds to the time of closest approach.}  
    \label{fig:coop}
\end{figure*}
\section{Results and Discussions}
\label{Scattering}

The collision or a scattering event in our system consists of two droplets approaching each other from a distance and their subsequent deviated motion when they feel each other's chemo-hydrodynamic fields. Here we present the experimental observations on the scattering of two self-propelling LC droplets. Care was taken so that the droplet suspension was not too dense to avoid three or more particle interactions.  The droplets with the same level of focus in the optical microscope and therefore residing in the same plane were considered 
for the collision studies.
 
The initial Peclet numbers of the droplets are given by, $Pe_{\alpha} = v_{\alpha}(0) R_{\alpha}/D_m$ where $v_{\alpha}(t)$ and $R_{\alpha}$ are the instantaneous speed and radius of the $\alpha$-th 
droplet respectively and the constant $D_m$ is the diffusivity of the filled micelles in the bulk micellar solution. The dimensionless Peclet number characterizes the relative magnitude 
of the advection and diffusion of the filled micelles. The dynamics of the two colliding droplets can be specified in a variety of ways. In our experiment without loss of generality, we 
denote the droplets with the lower and higher initial Peclet numbers as droplets 1 and 2, respectively. The initial and final velocities of the droplets are given by 
$\vect{v}_{\alpha}(0)$ and $\vect{v}_{\alpha}(t_f)$ ($\alpha = 1,2$), respectively, where $t_f$ is the duration of observation in our experiments.  

The diffusivity $D_m$ of filled micelles can be estimated as follows. The molecular size of 5CB is about 2 nm while the radius of spherical SDS micelles is reported to be 2.4 nm \cite{Todorov2002}. 
If the solubilization site is assumed to be at the center of the spherical micelles, we get an approximate radius for SDS-filled micelles to be $r_m \approx$ 3.4 nm. The filled micelle 
diffusivity can be determined from the Stokes-Einstein relation, $D_m =  k_BT/6\pi\eta r_m$, where $k_B$ is the Boltzmann constant, T is the absolute temperature and $\eta$ is the 
viscosity of the outer fluid. Taking $k_B  = 1.38\times 10^{-23} J \mhyphen K^{-1}$, $T = 313.15\,K$, $\eta  = 1 \times 10^{-3} \, Pa \mhyphen s $, we get 
$D_m \approx 6.74\times 10^{-11} \, m^2 \mhyphen s^{-1}$.

\begin{table*}[t]
  \centering
  \begin{tabular}{|c| c| c| c| c| c| c| c| c| c|}
  \hline
  
  $Pe_1$ & $Pe_2$   & $\chi(0)$    &  $r_{12}(0)(\mu m)$  & $\Delta \tau(s)$  & $\chi(t_f)$& $r_{12}^{min}(\mu m)$\\ [0.3ex]
  \hline
     2.3 & 3.4 & 0.731 & 105.10  & 5.13  & 0.96 &  52.5 \\ 
    \hline
     3.2 & 5.4 & 0.390 & 140.02 & 2.11 & 0.842 & 40.2  \\
    \hline
     3.7 & 4.2 & 0.574 & 92.67 & -1.52  & 0.87 &  44.3 \\
     \hline
     3.7 & 4.7 & 0.752 & 101.05 & -2.05  & 0.931 & 64.6  \\
    \hline
     3.2 & 4.4 & 0.705 & 73.61 &  2.55 & 0.97 & 43.4  \\
     \hline
     1.7 & 3.4 & 0.044  & 85.40 & 2.42  & 0.952  & 45.6  \\[1ex]
    \hline
  \end{tabular}
  \caption{The experimental collision parameter values for different sets of droplets corresponding to the acute angle of approach.}
  \label{tab:1}
\end{table*}
\begin{table*}[t]
  \centering
  \begin{tabular}{|c| c| c| c| c| c| c| c| c| c|}
  \hline
    $Pe_1$ & $Pe_2$   & $\chi(0)$    &  $r_{12}(0)(\mu m)$  & $\Delta \tau(s)$  & $\chi(t_f)$& $r_{12}^{min}(\mu m)$\\ [0.3ex]
  \hline
     3.7 & 4.3 & 0.988  & 67.6 & 1.141  & 0.976  & 40.4  \\ 
    \hline
     1.4 & 1.9 & 0.948  & 76.2 & 7.147  & 0.898  & 58.9   \\
    \hline
     2.5 & 3.2 & 0.984  & 68.3 & -2.094 & 0.987  & 59.7 \\
     \hline
     5.2 & 6.0 & 0.998  & 56.5 & 5.113  & 0.993  &  53.9 \\[1ex]
    \hline
  \end{tabular}
  \caption{The experimental collision parameter values for different sets of droplets corresponding to almost parallel angle of approach.}
  \label{tab:2}
\end{table*}
\begin{table*}[t]
  \centering
  \begin{tabular}{|c| c| c| c| c| c| c| c| c| c|}
  \hline
   $Pe_1$ & $Pe_2$   & $\chi(0)$    &  $r_{12}(0)(\mu m)$  & $\Delta \tau(s)$  & $\chi(t_f)$& $r_{12}^{min}(\mu m)$\\ [0.3ex]
  \hline
    3.6 & 4.9 & -0.957 & 82.8   & -13.1 & -0.972  &   31.5 \\ 
    \hline
    2.1 & 3.7 & -0.991 & 71.4   & -30.1  & -0.970 &   31.4\\
    \hline
    2.1 &3.7  & -0.985 & 130.1  & 21.3    & -0.924 &   31.4 \\
    \hline
    4.0 & 4.2 & -0.971 & 125.4  & -7.0    & -0.973 &   38.3 \\
    \hline
    1.5 & 3.1 & -0.986 & 108.7 & -8.9   & -0.99  &   38.1 \\
    \hline
    1.7 & 4.1 & -0.900 & 72.5   & 15.5   & -0.986 &   41.9\\
    \hline
    0.9 & 3.3 & -0.970 & 94.1   & -11.8   & -0.948 &   36.4\\
    \hline
    1.6 & 2.0 & -0.784 & 76.2   & -6.7    & -0.990 &   35.1\\
    \hline
    2.0 & 2.6 & -0.754 & 97.1   & -8.1    & -0.960 &   43.6\\
    \hline
    1.9 & 3.6 & -0.935 & 150.9 & -17.9  &  -0.99 &   47.2\\
    \hline
    0.9 & 1.9 & -0.996 & 138.0 & 72.4   & -0.998 &   39.1\\
    \hline
    1.0 & 2.3  & -0.997  & 97.2 & 73.9   & -0.990 &   39.9\\
    \hline
    0.7& 3.1 & -0.871  & 125.5  & 12.8    & -0.964 &   33.4\\[1ex]
    \hline
  \end{tabular}
  \caption{The experimental collision parameter values for different sets of droplets corresponding to almost anti-parallel angle of approach.}
  \label{tab:3}
\end{table*}
\begin{table*}[t]
  \centering
  \begin{tabular}{|c| c| c| c| c| c| c| c| c| c|}
  \hline
 $Pe_1$ & $Pe_2$   & $\chi(0)$    &  $r_{12}(0)(\mu m)$  & $\Delta \tau(s)$  & $\chi(t_f)$& $r_{12}^{min}(\mu m)$\\ [0.3ex]
  \hline
    2.6 & 3.3 & -0.237  & 128.0 & 5.3     &  0.998  & 47.5 \\ 
    \hline
    1.0 & 1.8 & -0.530  & 70.3  & 3.2    & 0.971   &  25.3 \\
    \hline
     3.4 & 4.7 & -0.974 & 164.7  & -5.4   &  0.995  &  31.3 \\
    \hline
    1.5 & 2.2 & -0.697  & 79.1   &  -6.4   &  0.990  &  40.0 \\
    \hline
    1.1  & 1.3 & -0.192 & 124.9  &  9.0    & 0.935   & 43.1 \\
    \hline
    4.1 & 5.0 & -0.281  & 172.5  & -4.2    &  0.925  &  48.3 \\
    \hline
    2.1 & 3.5 & -0.521  & 98.3   & -0.3   &  0.929  & 41.9  \\
    \hline
    0.6 & 1.2 & -0.458  & 84.6   &  5.8    &  0.880  &  35.2 \\
    \hline
    1.2 & 2.1 & -0.686 & 79.7    & 2.3     &  0.816  &  41.2\\
     \hline
    3.5 & 3.5 & -0.806 & 110.0  & 1.6     & 0.926   &  41.6\\
     \hline
    1.6 & 4.4 & -0.233 & 72.9  & 2.1    &  0.926  &  41.8\\
     \hline
   4.3  & 8.4 & -0.121 & 97.4    &  2.6   &  0.976  & 45.9 \\
     \hline
    1.5 & 2.4 & -0.232 & 89.8    &  -0.7 &  0.743  &  36.0\\
    \hline
    0.8 & 3.1 & -0.162 & 115.9   & -6.0   &  0.932  &   45.1\\[1ex]
    \hline
  \end{tabular}
  \caption{The experimental collision parameter values for different sets of droplets corresponding to an obtuse converging angle of approach.}
  \label{tab:4}
\end{table*}

 The dynamical state of the droplets during the scattering event at time $t$ can be represented by the relative direction cosines of their instantaneous velocities defined as \cite{Lippera2021Alignment},
\begin{equation}
 \chi (t) = \frac{\vect{v}_1(t) \cdot \vect{v}_2(t)}{|\vect{v}_1(t)|  |\vect{v}_2(t)|}   
\end{equation}
Thus $\chi(0)$ and $\chi(t_f)$ denote the initial and final relative direction cosines of the droplets, respectively.  
Following Lippera et al. \cite{Lippera2021Alignment}, we define the point at which the directions of initial velocities intersect as the virtual crossing point (VCP). The distance of particle $\alpha$ from the virtual crossing point at instant $t$ is given by, 
  \begin{equation}
  l_{\alpha}(t) = [(x_{vcp} - x_{\alpha}(t))^2 + (y_{vcp} - y_{\alpha}(t))^2]^{1/2} ,\,
 \end{equation}
 where, ($x_{vcp}$, $y_{vcp}$) is the coordinates of the VCP and $(x_{\alpha}(t), y_{\alpha}(t))$ is the instantaneous position of the $\alpha$-th particle at time $t$. 
In their theoretical study, Lippera \textit{et al.} considered $(l_1(0)-l_2(0))$ as one of the initial conditions for their scattering events.  However, they considered symmetric and asymmetric 
collisions between equal-sized droplets with equal speeds for arbitrary relative initial angles \cite{Lippera2021Alignment}. It is highly improbable to find equal-sized droplets and 
symmetric collisions in our experimental setting. In general, the droplets are of unequal sizes and they do not attempt to reach the virtual crossing point symmetrically. In some cases, 
the smaller droplet is lagging while in some other cases, the larger droplet is lagging to reach the virtual crossing point. Therefore, to quantify the initial conditions of the droplets 
in our experimental studies, we define the virtual time  $\tau_{\alpha} = l_{\alpha}(0)/v_{\alpha}(0)$. We further consider $\tau_{\alpha}$ as a signed quantity that can be positive or 
negative depending on the $\alpha$-th particle approaching towards or receding from the virtual crossing point at $t=0$, respectively. Therefore in the case of the droplet approaching 
to or receding from the virtual crossing point, $dl_{\alpha}/dt|_0 $ is less than or greater than zero. We define the virtual lag time  $\Delta \tau  = \tau_1 - \tau_2$ as one of the 
initial conditions in our experimental studies. We also determine the inter-particle distance during the scattering event as given by,
\begin{equation}
    r_{12}(t) =[(x_1(t)-x_2(t))^2+(y_1(t)-y_2(t))^2]^{1/2}.
\end{equation} 
Thus the initial conditions of the colliding droplets in our experimental studies can be specified by the following set of parameters  \{$Pe_1, Pe_2,r_{12}(0), \Delta \tau, \chi(0)$\}. 
We experimentally follow the variation of $\chi(t)$, $v_{\alpha}(t)$, and $r_{12}(t)$ as a  function of time.

The larger droplets have a greater tendency to turn during their self-propulsive motion than smaller droplets \cite{Suda2021, Ray2023Experimental}. The change in direction of motion can arise 
due to two reasons, namely, the effect of another droplet in the vicinity of the moving droplet and its secondary instability at the interface. However, even for larger droplets, their 
motion is ballistic at sufficiently small time scales which allows us to define the velocities of the droplets.

The experimental studies were carried out for different sets of droplets with different initial conditions. Based on our experimental observations, we identify four types of scattering events as shown 
in Figure~\ref{fig:collision}. 


 (a) \textbf{Acute angle of approach}: Droplets with an acute angle of approach and small delay time $\Delta \tau$ have a more acute course of motion after the collision as shown in Fig.~\ref{fig:collision}(a). This general trend of the final angle being less than the initial angle is observed for converging trajectories. Such type of scattering is observed for $\Delta \tau$ in the range of about $-2.1 \, s$ to $ 5.1 \,s$ in our experimental studies. This type of aligning effect has also been found in some theoretical models \cite{Lippera2021Alignment, Yabunaka2016}.    Fig.~\ref{fig:collision}(b) shows the corresponding variation of the relative direction cosine $\chi$ and inter-particle distance $r_{12}$ of the droplets as a function of time. The minimum in the $r_{12}$ corresponding to the value $r^{min}_{12}$ $\approx$ 44 $\mu m$ represents the closest approach of the droplets during the collision. The parameter $\chi$ changes from an acute value of about 0.17 to nearly 1 after the collision confirming the aligning effect. Fig.~\ref{fig:collision}(c) shows the variation of the speed of the droplets as a function of time. In this type of collision, it is found that the initial and the final speeds of the droplets do not vary significantly. Several experimental studies for similar scattering events have been carried out with the parameter values as listed in Table ~\ref{tab:1}.

 (b) \textbf{Almost parallel approach}: The droplets moving almost parallelly before the collision were found to remain almost parallel after the collision. This is observed for both equal and slightly unequal-sized particles. Fig.~\ref{fig:collision}(d) shows the trajectories of two droplets undergoing such a collision event.  This type of scattering is observed for delay time $\Delta \tau$ in the range $-2.1 \, s$ to $ 22.4 \, s$. Fig.~\ref{fig:collision}(e) shows the variation of the relative direction cosine $\chi$ and inter-particle distance $r_{12}$ of the droplets as a function of time.  The relative direction cosine remains close to 1 for such a collision over the entire duration of observation. The distance of closest approach is $r^{min}_{12}$ $\approx$ 74 $\mu m$ in this case. Fig.~\ref{fig:collision}(f) shows the variation of the speed of the droplets as a function of time.  The parameter values corresponding to a few of these collision events are listed in Table~\ref{tab:2}.

(c) \textbf{Almost antiparallel approach}: In this type of collision, droplets moving with an almost antiparallel angle of approach tend to recede each other in an antiparallel orientation after the collision as shown in Fig.~\ref{fig:collision}(g).  Droplets with initial direction cosine $-0.75\lesssim \chi(0)\lesssim -1.0$ and delay time $-30.1\, s \lesssim \Delta \tau \lesssim 73.9 \,s$ showed this kind of collision. This kind of collision event was also found in a recent theoretical study \cite{Lippera2021Alignment}.   Fig.~\ref{fig:collision}(h) shows the variation of the relative direction cosine $\chi$ and inter-particle distance $r_{12}$ of the droplets as a function of time. The minimum value of inter-particle distance is $r_{12}^{min}$ $\approx$  39 $\mu m$. Fig.~\ref{fig:collision} (i) shows the speeds of these droplets as a function of time. The speed of the respective droplets remains almost the same before and after the collision which can be seen from Fig.~\ref{fig:collision}(i).  Similar collision events were studied for several droplet pairs in our experiments.  The parameter values corresponding to a few of these collision events are listed in Table~\ref{tab:3}.

 (d) \textbf{Obtuse angle of approach}:  Droplets moving with a converging obtuse angle of approach also showed an aligning interaction as shown in Fig.~\ref{fig:collision}(j).  The variations of the relative direction cosine and the interparticle distance between the droplets during the collision event are shown in Fig.~\ref{fig:collision}(k). For this representative collision, the initial angle between the velocities is close to $\pi$ i.e. $\chi(0)$ is close to -1, and the final angle is closer to $0$, meaning the relative angle has decreased. The minimum inter-particle distance is $r^{min}_{12}$ $\approx$ 33 $\mu m$. The point of closest approach roughly corresponds to the point of maximum direction change. The corresponding change in speed is shown in Fig.~\ref{fig:collision}(l). Several experimental studies for similar scattering events have been carried out with the parameter values as listed in Table ~\ref{tab:4}.

Ideally, when the droplets are sufficiently separated before and after the collision, their motions are mostly independent and it is expected that for a steady dissolution rate of the droplets,  the final speed should be similar to the initial speed for each droplet. This is observed in most of our experiments except for the motion of droplet-2 as shown in Fig.~\ref{fig:collision}(f). In this case, droplet-2 initially has a relatively larger speed than droplet-1 but after collision, they tend to move with similar speeds. It perhaps arises because the droplets did not sufficiently separate after the collision due to their almost parallel motion.

In some previous studies, a transient bound state between two droplets was observed to form for some initial conditions of the droplets \cite{Hokmabad2022}. In our experiment, it was found that the droplets formed a transient bound state after the collision for the initial Peclet number in the range of about 8 to 20. Two droplets with Peclet numbers about 11 and 20 were found to move in a cooperative manner (forming a transient bound state) if their initial courses of motion were converging as shown in Figure \ref{fig:coop}(a). Figure  ~\ref{fig:coop}(b) shows the variation of $\chi$ and $r_{12}$ as a function of time during the collision of these two droplets. After the collision, the relative direction cosine $\chi$ saturates to 1 and the separation $r_{12}$ becomes almost constant during the observation period implying that the droplets move in a compact fashion.  The variation in the speed of the droplets is shown in Fig. ~\ref{fig:coop}(c).
 The bigger droplet has a relatively higher speed than the smaller one implying that the bound state is transient in nature. As stated earlier, the Peclet number represents the ratio of advection and diffusion of the filled micelles. For higher Peclet numbers, advection is more dominant than diffusion around the droplet. Therefore, for collision between two small droplets the chemical field is expected to be dominant whereas for collision between two large droplets, hydrodynamic interaction is expected to be dominant. The orbiting motion of squirmers predicted theoretically \cite{Darveniza2022} was not observed in our droplet system.


\section{Conclusion}
\label{Discussion}
The generic planar collisions of two droplets relevant to experimental conditions were studied. Swimming droplets influence each other’s dynamics through
the wake of chemical solutes they generate to self-propel themselves and through their hydrodynamic fields. The scattering dynamics of the droplets are governed by a combination of far-field hydrodynamic attraction, and near-field chemical repulsion.   In our experiment, we identify four different kinds of frequently occurring oblique collisions. Droplets with sufficiently high Peclet numbers were seen to form transient bound states for converging angles of approach. The maximum reorientation in the motion of the droplets during collision was found to occur near the point of closest approach.

\newpage
\bibliographystyle{apsrev4-2.bst}
\bibliography{references.bib}

\end{document}